\documentclass{aa}   

\usepackage{amssymb} 
\usepackage{multirow}
\usepackage{graphicx}
\usepackage{amsmath}
\usepackage[normalem]{ulem}
\usepackage{booktabs}
\usepackage{tablefootnote} 
\usepackage{txfonts}
\useunder{\uline}{\ul}{}
\usepackage{ulem} 
\usepackage{float}
\usepackage{hyperref}
\usepackage{bm}
\usepackage{threeparttable}
\usepackage{aas_macros}
\usepackage{subcaption}
\renewcommand\thesubfigure{\roman{subfigure}}
\begin{document} 

   \title{Dust shells and dark linear structures on dust tails of historical and recent long-period comets}

    \author{Fernando Moreno \inst{1} 
\and
    Emmanuel Jehin \inst{2}
         } 
   \institute{Instituto de Astrof\'{i}sica de Andaluc\'{i}a, CSIC, Glorieta de la Astronom\'{i}a s/n, E-18008 Granada, Spain \\
              \email{fernando@iaa.es}
\and
STARInstitute, Université de Liège, Allée du 6 Août 19c, 4000 Liège, Belgium}              

\titlerunning{Shells and dark linear structures on long-period comets}  
 \authorrunning{Fernando Moreno}
 \abstract
   {Dust halos or shells, along with linear dark structures along the axes of dust tails, are commonly observed in many long-period comets near perihelion. Examples range from the recent C/2023 A3 (Tsuchinshan-ATLAS) to historical comets such as the Great Comet of 1874, C/1874 H1 (Coggia).}
   {While dust halos can readily be modeled as spin-modulated activity originating from the comet nucleus, their possible connection to those dark linear features has, to our knowledge, not been investigated. The aim of this paper is to shed light on the formation of these remarkable structures by modeling a sample of six long-period comets, using similar dust physical properties and ejection parameters, to explore whether they share a common origin.}
   {To model the dust features, we employed a Monte Carlo procedure to generate synthetic images. The particles ejected from the comet nucleus follow a power-law size distribution and are released into interplanetary space at speeds determined by the ratio of solar radiation pressure to solar gravity, the heliocentric distance, and, as a new feature of the code, the solar zenith angle at the emission point.}
  {We demonstrate that, in all the cases analyzed, the dust shells form as a result of short-term events characterized by cyclically varying ejection of very small particles from large surface areas on the rotating nucleus. These events are triggered as these areas become freshly exposed to solar radiation near perihelion due to the high obliquity of the spin axes of their nuclei. The dark linear stripes along the tail axes may arise from a specific dependence of the ejection speeds on the square root of the cosine of the zenith angle, as is predicted by hydrodynamical modeling, but 
  their presence is also dependent on the extent of the latitude region of emission that defines the velocity vector field.}
  {}
  
   \maketitle
%

\section{Introduction} \label{sec:intro}

\begin{figure}
\centering\includegraphics[trim={2.5cm 2.5cm 3cm 0cm}, clip,angle=-90,width=\columnwidth]{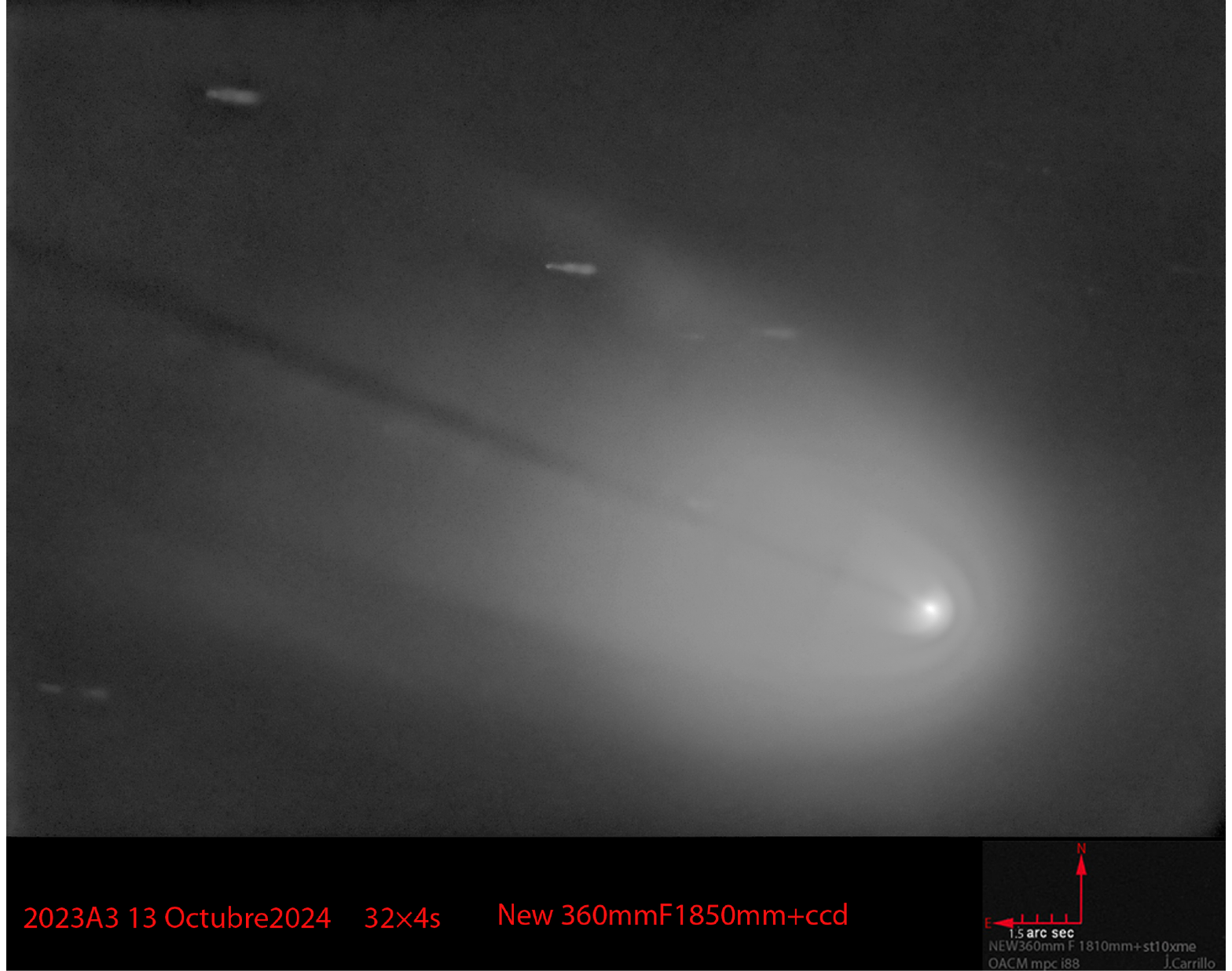}
\caption{Image of C/2023 A3 (Tsuchinshan-ATLAS) obtained on October 13, 2024, by amateur observer Jos\'e Carrillo. Dust shells and a DLS along the tail axis are clearly seen. North is up; east is to the left.}
\label{fig:C2023A3-1}
\end{figure}

Recent high-spatial-resolution amateur images of the Great Comet of 2024, C/2023 A3 (Tsuchinshan-ATLAS), have revealed a remarkable level of detail in its dust structures, particularly the presence of sunward dust shells or halos and a striking dark linear structure (hereafter DLS) along the tail axis (see Fig. \ref{fig:C2023A3-1}). Through a search of past long-period comets, we have identified similar structures in several cases, as we shall demonstrate in the following sections.

Dust halos and jets have been extensively explained by the presence of active sources on rotating, spherically shaped nuclei, located at specific latitude-longitude positions
\citep[e.g.,][and references therein]{1998ApJ...509L.133S, 2004ApJ...616.1278V}. In these studies, specific choices of particle sizes and ejection times were necessary to account for the variety of features observed in the tails. A significant advancement in interpreting such features was made by \cite{2010A&A...512A..60V}, who applied a similar approach to comet 9P/Tempel 1 while incorporating observations from the Deep Impact mission regarding nucleus shape, illumination conditions, and other relevant parameters.

In most cases, including the objects described here, no direct information on the nuclei is available. Therefore, we must rely on the simplest model -- a rotating spherical nucleus with an active area on it. We refer to this model as to the single active area model. Our objective is to demonstrate that the remarkable structures observed in C/2023 A3 are common among long-period comets near perihelion and that they can be explained using similar dust physical parameters, speed distributions, and spin axis orientations in all cases.

\section{Description of the long-period comets selected} \label{sec:observations}

\begin{figure}
\centering
\centering\includegraphics[trim=1cm 2cm 3cm 3cm,clip,
width=0.5\textwidth]{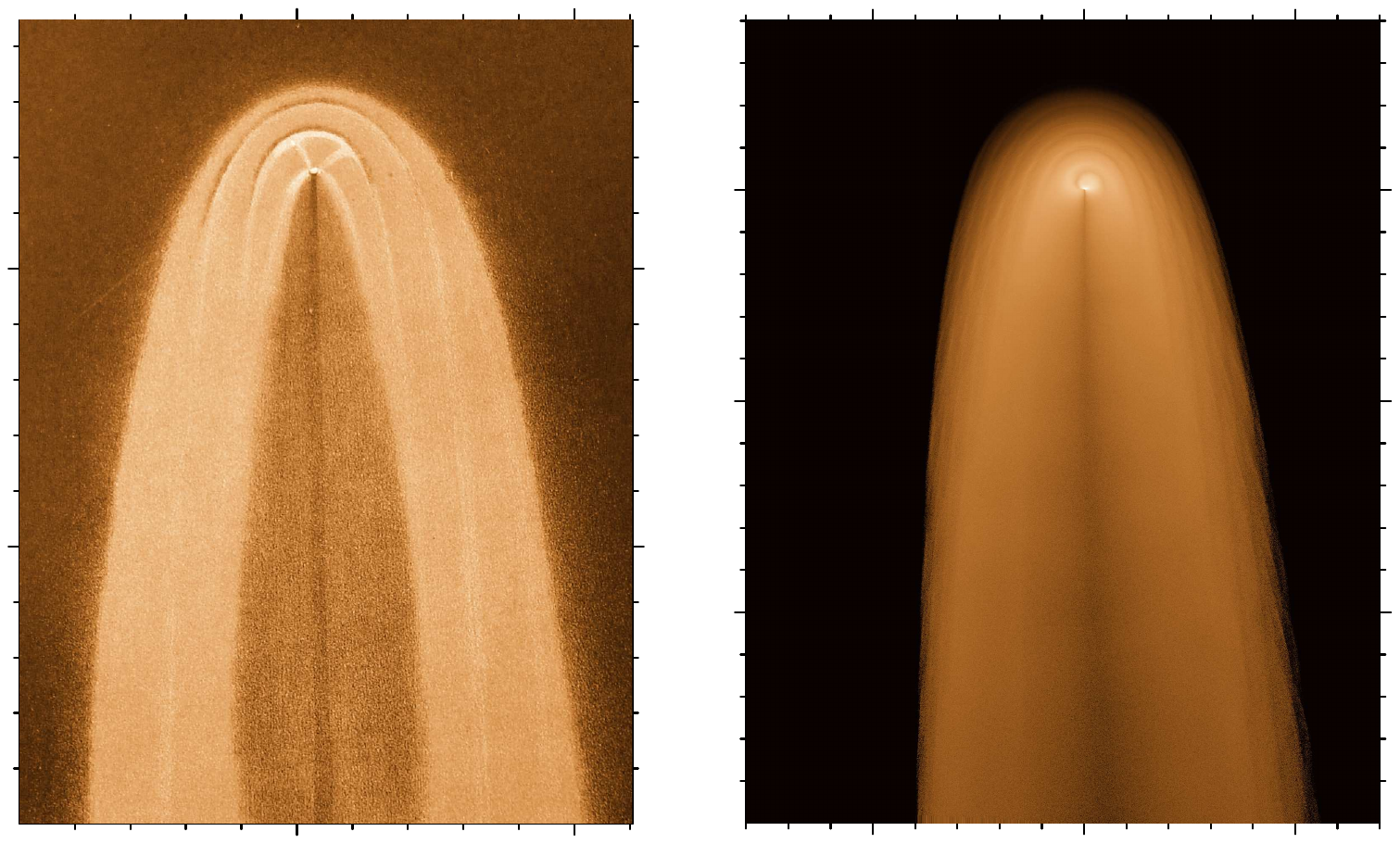}
\caption{Comet C/1858 L1 (Donati). The left panel is from Wikipedia\protect\footnotemark.  The engraving was created by W.C. Bond and is featured in the book Astronomie by Karl von Littrow, published in 1878. It is in the public domain, via Wikimedia Commons. The right panel is the simulated image. The image dimensions are  616,850$\times$781,343 km projected on the sky.}
\label{fig:C1858L1}
\end{figure}

\begin{figure}
\centering\includegraphics[trim=1cm 3cm 2cm 4cm,clip,width=0.5\textwidth]{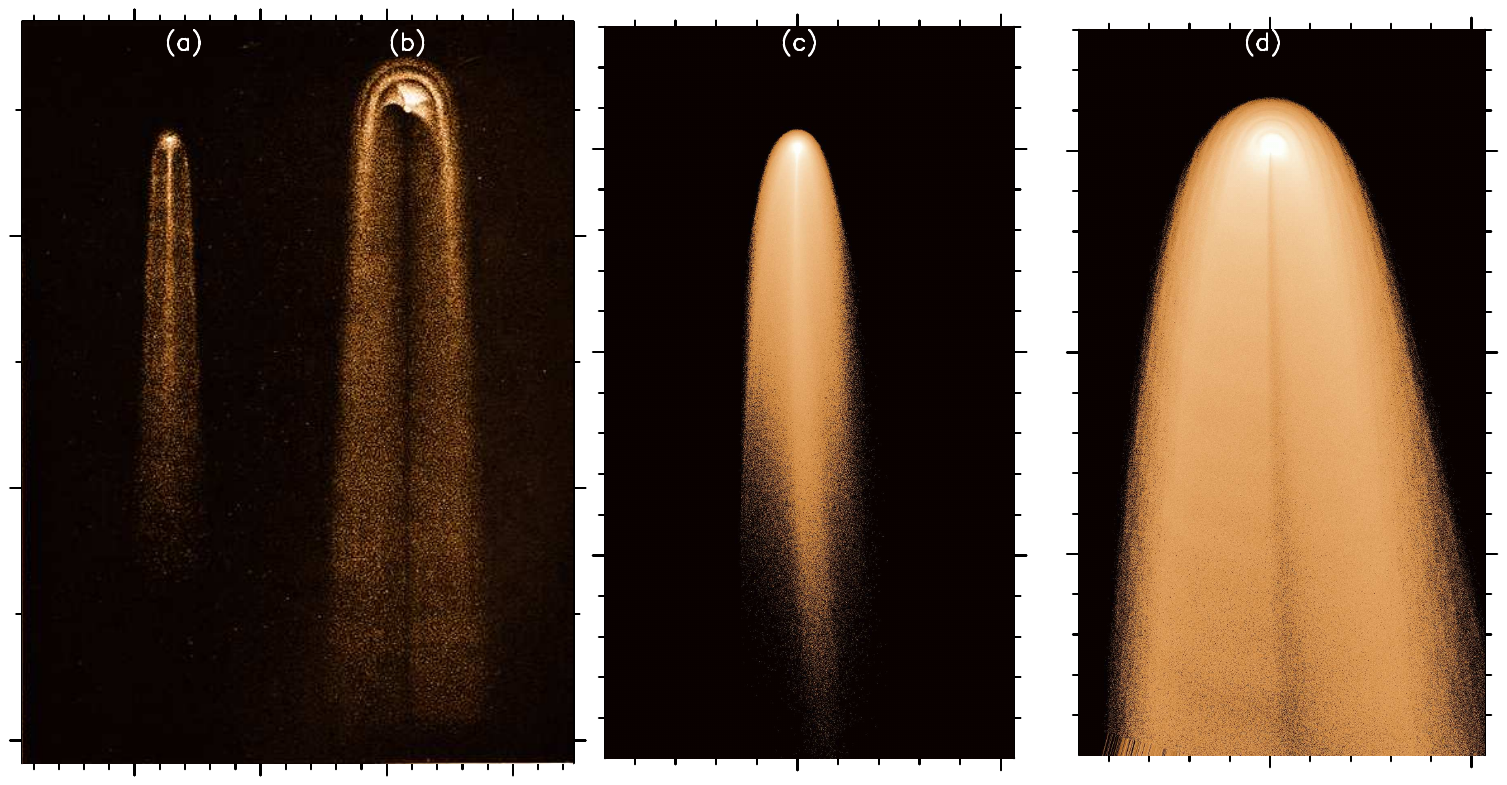}
\caption{Comet C/1874 H1 (Coggia). The leftmost panel shows the observations on June 10, 1874 (a), and July 9, 1874 (b). Central panel (c) and rightmost panel (d) show the simulations corresponding to images (a) and (b), respectively. In panel (c) the image dimensions are  3,144,105$\times$5,620,050 km, and in panel (d) they are 614,693$\times$1,099,800 km projected on the sky. The figure in the leftmost panel is from Wikipedia\protect\footnotemark, by 
Robert Stawell Ball (1840-1913). It is in the public domain, via Wikimedia Commons.}
\label{fig:C1874H1}
\end{figure}

\begin{figure} 
\centering\includegraphics[trim=1cm 1.5cm 4cm 2.5cm,clip,
width=0.5\textwidth]{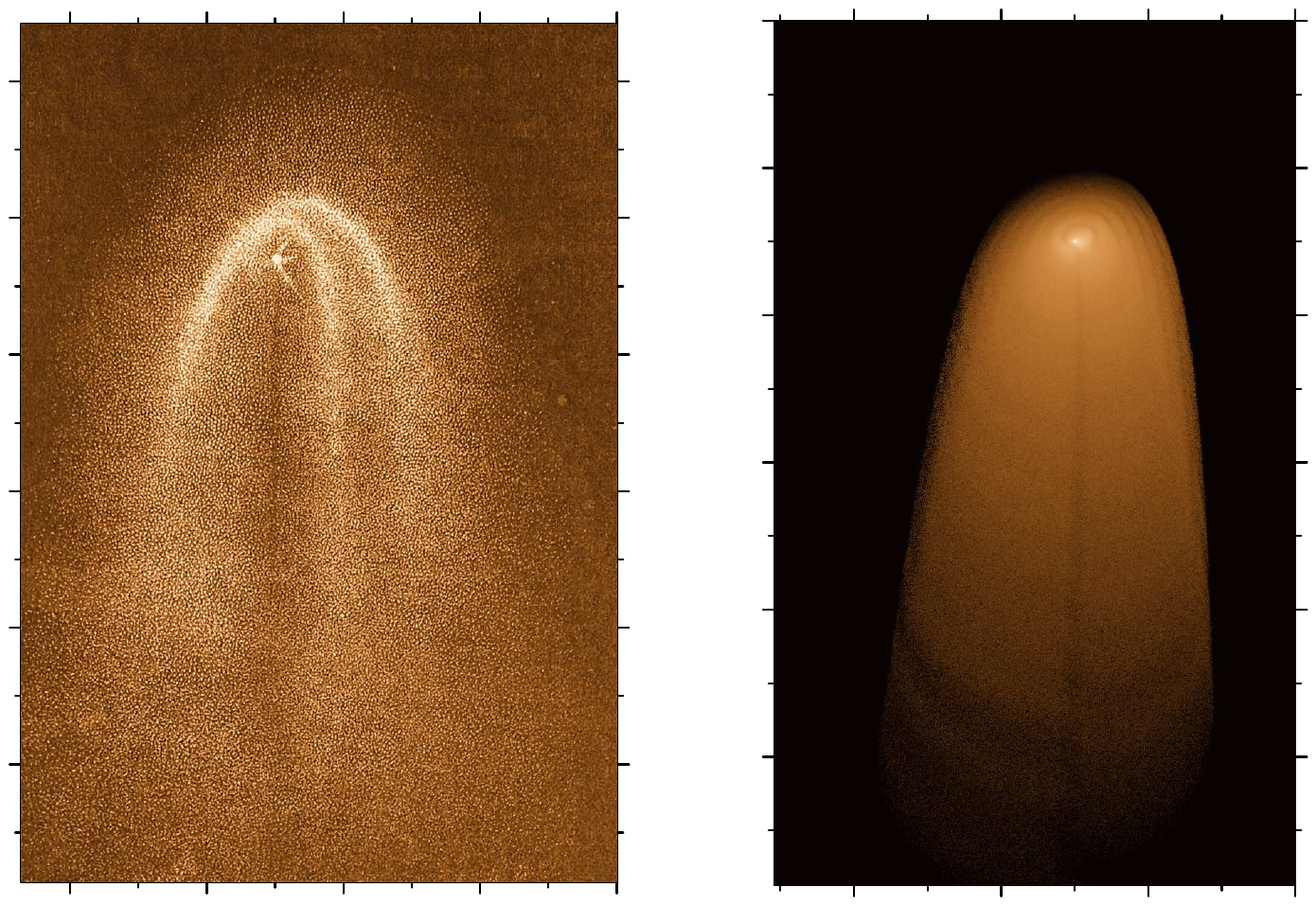}
\caption{Comet C/1881 K1 (Tebbutt). The left panel is from 
Wikipedia\protect\footnotemark. 
The comet was observed and drawn by Dr. L. Weinek. The image is in the public domain, via Wikimedia Commons.
The simulated image in the right panel has dimensions of 667,545$\times$1,106,918 km projected on the sky.
}
\label{fig:C1881K1}
\end{figure}

\begin{figure}
\centering\includegraphics[trim=1cm 4cm 5cm 5cm,clip,
width=0.5\textwidth]{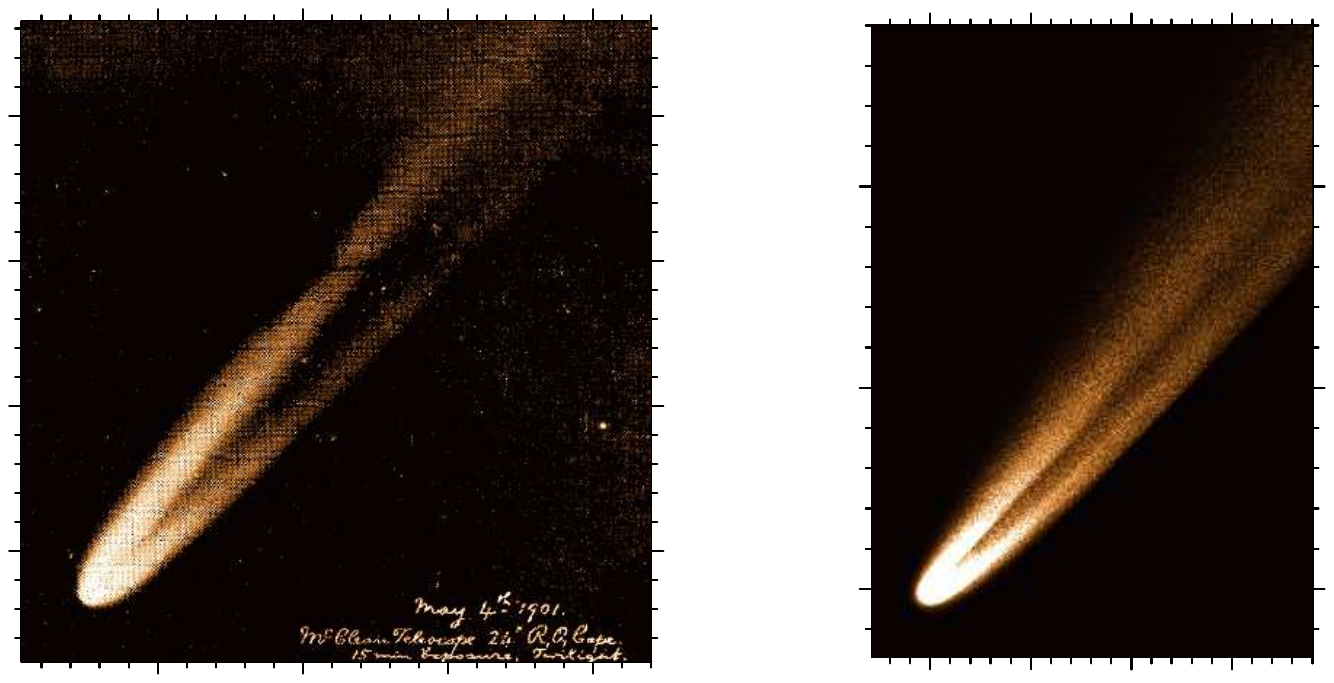}
\caption{Comet C/1901 G1 (Viscara). The left panel is from 
Wikipedia\protect\footnotemark, a photograph obtained by Edward Emerson Barnard, 1857-1923, during the United States Naval Observatory expedition to Sumatra. It is in the public domain, via Wikimedia Commons. The simulated image in the right panel has dimensions of 735,378$\times$982,689 km projected on the sky.
}
\label{fig:C1901G1}
\end{figure}

\begin{figure*}
\centering\includegraphics[trim=0.5cm 4cm 4cm 6cm,clip,width=\textwidth]{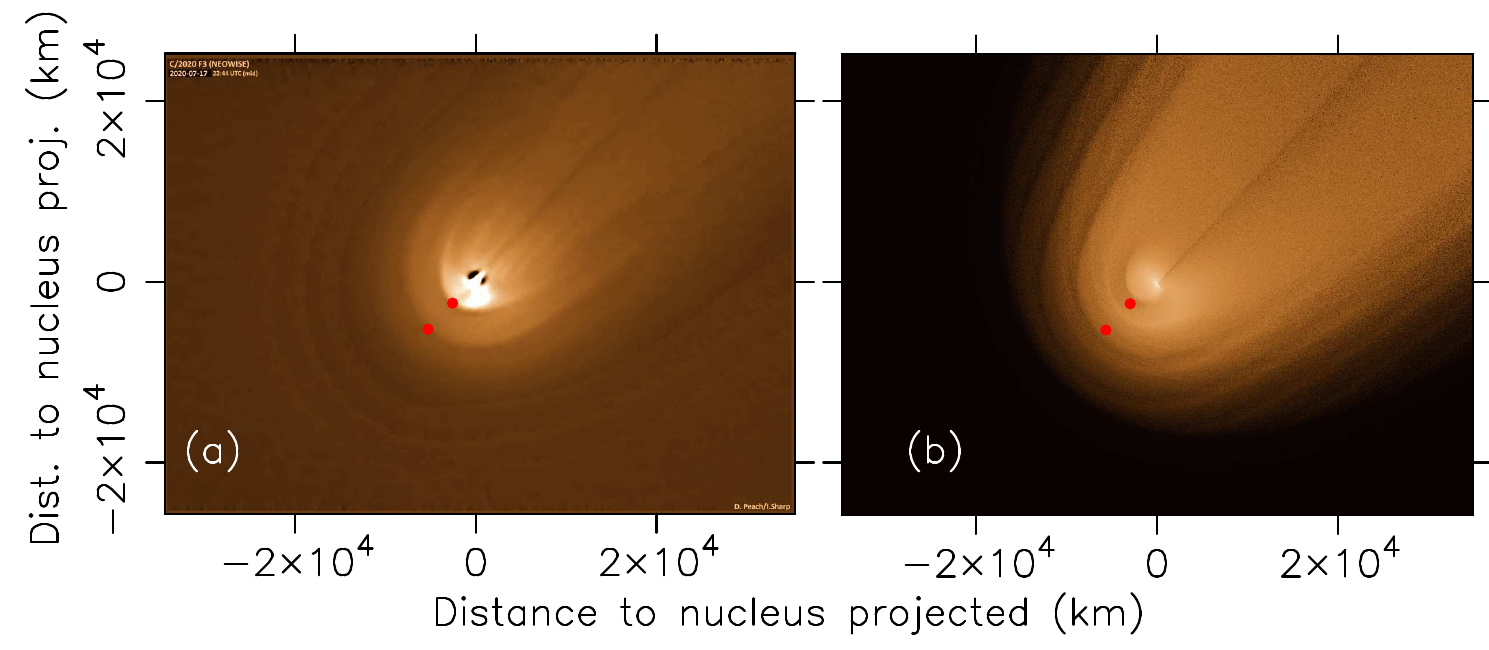}
\caption{Comet C/2020 F3 (NEOWISE). Panel (a) is an image obtained from Damian Peach\protect\footnotemark, reprinted with the author's written permission. 
Panel (b) is the simulated image. This synthetic image is rotated by 66$^\circ$ westward with respect to the conventional north-up, east-to-left orientation to approximately match the orientation of the image in panel (a). Red dots indicate the position of the first and second dark intervals between shells along the tail axis in both the observation and the simulation.}
\label{fig:C2020F3}
\end{figure*}

\begin{figure*}
\centering\includegraphics[
trim=0cm 2.5cm 3cm 4.3cm,clip,width=\textwidth]{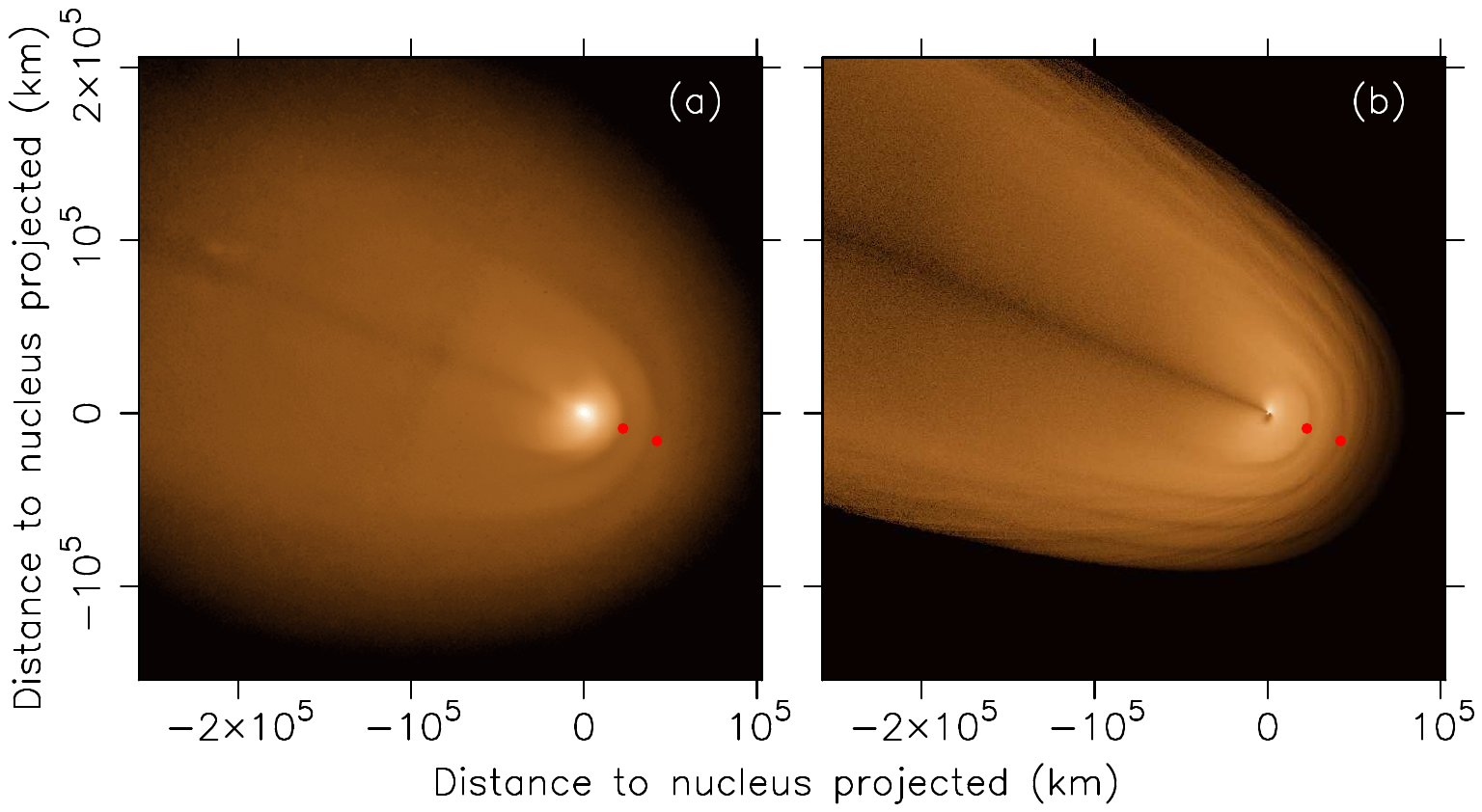}
\caption{Comet C/2023 A3 (Tsuchinshan-ATLAS). Panel (a) is an image obtained by Jos\'e Carrillo\protect\footnotemark, reprinted with the author's written permission. Panel (b) is the simulated image.  Red dots indicate the position of the first and second dark intervals between shells along the tail axis in both the observation and the simulation. North is up; east is to the left.}
\label{fig:C2023A3}
\end{figure*}

We have selected a subset of long-period comets that exhibit similar morphology near perihelion, specifically dust shells and a DLS along the tail axis. Table \ref{tab:logobs} presents the orbital elements of the selected objects, while Table \ref{tab:modelpar} provides the observation dates when these features were recorded, along with the model parameters used to reproduce them, which will be discussed in detail in Sect. \ref{sec:model}. Images of each comet are displayed in Figs. \ref{fig:C1858L1} to \ref{fig:C2023A3}.

For observations prior to 1900, the representations consist of either drawings or engravings made by various observers, as is indicated in the figure captions. The comet C/1901 G1 was captured on a photographic plate, whereas the two most recent comets, C/2020 F3 and C/2023 A3, were imaged using modern CCD detectors. All selected comets have perihelion distances well within 1 au and exhibit a range of orbital inclinations, following either prograde or retrograde orbits around the Sun. Concerning the DLS, we note that a short perihelion distance is not a necessary requirement for the DLS to appear. Although we are not aware of any such cases, it is possible that the DLS could also appear in comets with larger perihelion distances. In all cases, except for comet C/1901 G1 -- possibly due to a relatively large field of view -- the dust shell structures are clearly visible, as is the DLS along the axis of the dust tails. 
 In all cases, the mentioned structures were observed at perihelion or later (see Table \ref{tab:modelpar}), with the exception of the pre-perihelion observation of comet C/1874 H1 (Coggia), which does not exhibit these features. This particular observation is modeled as being characterized by a sunward particle emission pattern, as will be described in Sect. \ref{sec:results}.

\begin{table*}
  \centering
  \caption{Ecliptic osculating orbital elements of the different objects.$^\star$ }   
  \label{tab:logobs}
 
  \begin{tabular}{|l|l|l|l|r|r|r|l|}
    \hline
\multicolumn{1}{|c|}{Object} & \multicolumn{1}{c|}{Epoch (JD)} & \multicolumn{1}{c|}{$e$} & 
\multicolumn{1}{c|}{$q$(au)}& \multicolumn{1}{c|}{$\Omega$($^\circ$)} & \multicolumn{1}{c|}{$w$($^\circ$)}    & \multicolumn{1}{c|}{$i$($^\circ$)}& \multicolumn{1}{c|}{T$_p$(UT)}  \\
    \hline
C/1858 L1 (Donati) & 2399960.5 & 0.99629 & 0.578 & 167.30 & 129.11& 116.95 & 1858-Sep-30.46  \\
C/1874 H1 (Coggia) &2405720.5  &0.99820 & 0.676 & 120.49 & 152.38 & 66.34 & 1874-Jul-09.35   \\
C/1881 K1 (Tebbutt) &2408306.5 & 0.99590 & 0.734 & 272.63 & 354.23 & 63.43 & 1881-Jun-16.94  \\
C/1901 G1 (Viscara) & 2415523.5 & 1.00000 & 0.245 & 111.04 & 203.05 & 131.08 & 1901-Apr-24.75 \\   
C/2020 F3 (NEOWISE) &2459036.5 & 0.99918 & 0.295 & 61.01 & 37.28 & 128.94 & 2020-Jul-03.68 \\
C/2023 A3 (Tsuchishan-ATLAS) &2460342.5 & 1.00012 & 0.391 & 21.56 & 308.49 & 139.11 & 2024-Sep-27.74  \\
  \hline

\multicolumn{8}{l}{$^\star$$e$:eccentricity; $q$:perihelion distance; $\Omega$:longitude of ascending node; $w$:argument of perihelion; $i$:inclination; T$_p$:perihelion time.} \\  
  
  \end{tabular}

  
  \end{table*}

\footnotetext{\url{https://commons.wikimedia.org/wiki/File:Comet_Donati_by_Bond_1858.jpg}}

\footnotetext{\url{https://commons.wikimedia.org/wiki/File:Comet_Coggia,_1874.jpg}}

\footnotetext{\url{https://commons.wikimedia.org/wiki/File:Die_Gartenlaube_(1881)_b_501_1.jpg}}

\footnotetext{\url{https://commons.wikimedia.org/wiki/File:Great_comet_of_1901.jpg}}

\footnotetext{\url{https://www.damianpeach.com/c2020f3.htm}}

\footnotetext{\url{http://www.astrosurf.com/cometas-obs/}}

\section{The Model} \label{sec:model}

We used a Monte Carlo dust tail code to simulate the observed images. The numerical code has already been described in several papers, and the most recent version of the program, available on GitHub\footnote{\url{https://github.com/FernandoMorenoDanvila/COMTAILS/}},  
is described in \cite{Moreno25a}. The code works under the assumption that the ejected dust particles are only affected by solar gravity and solar radiation pressure, ignoring the nucleus gravity and the hydrodynamic gas drag, an assumption that should be valid for small size nuclei, at a distance of $\sim$20 nuclear radii, describing Keplerian trajectories around the Sun. The particles are distributed in size by a power-law function, and their speed is a function of the $\beta$ parameter (i.e., the ratio of the solar pressure force to the solar gravity force), the comet heliocentric distance, and the cosine of the solar zenith angle at the ejection point. Both the nucleus and the particles are assumed to be spherical, and the ejection occurs in the nucleus radial direction. The code works under three different ejection regimes; namely, isotropic, hemispherical, or restricted to some “active areas.” It is assumed that the particles do not experience any mass loss (e.g., evaporation or disruption processes) or density change, and that they move in a collisionless regime. In all cases, the emission is assumed to occur uniformly across the sunlit portion of the nucleus. Additionally, since the described structures are typically observed for only a few days near perihelion, we assume that the activity responsible for generating these features lasts only a few days before the observation time. This transient activity would be superimposed on the background cometary dust emission, which generally originates from larger areas of the nucleus surface than those required to produce those peculiar structures.

The $\beta$ parameter can be written as 
$\beta = \frac{C_{pr} Q_{pr}}{2\rho r}$ \citep{1968ApJ...154..327F}, where $C_{pr} = 1.191 \times 10^{3}$ kg m$^{-2}$ is the radiation pressure coefficient, $Q_{pr}$ is the scattering efficiency for radiation pressure, $\rho$ is the particle density, and $r$ is the particle radius. 
For moderately absorbing particles of sizes larger than $r \gtrsim$ 1 $\mu$m at incident red wavelengths ($\lambda\sim$0.65 $\mu$m),  Mie theory predicts $Q_{pr}\sim$1
\citep[see e.g.][their Fig. 5]{2012ApJ...752..136M}. However, for particles smaller than the wavelength of the incident light, $Q_{pr}$ depends significantly on the particle radius and their complex refractive index, $n$. Since the modeling of the dust structures described below requires the presence of small particles, we need to provide the corresponding function, $Q_{pr}(n,r)$, which we calculate with Mie theory. In addition, the geometric albedo of the particles is also dependent on those two parameters. The geometric albedo can be defined as $p_v(\alpha)=\pi S_{11}(\alpha)/(G k^2)$, where $S_{11}(\alpha)$ is the first row and first column element of the scattering matrix, $\alpha$ is the phase angle, $k=2\pi/\lambda$, where $\lambda$ is the effective wavelength of the band filter used, and $G$ is the geometrical cross section of the particle, $G=\pi r^2$. The geometric albedo is also dependent on the particle composition (the refractive index) through $S_{11}(\alpha)$, which is computed with Mie theory as well. To represent a typical cometary composition of a mixture of silicate, carbon, and organic material, the Mie calculations were performed assuming a refractive index of $n$=1.6+0.01$i$, which falls within the range of refractive indices assumed for cometary dust  \citep{2020Icar..33613453Z}, although these values may vary from comet to comet.

Particle speeds were parameterized following common practice, depending on size, heliocentric distance ($r_h$), and the cosine of the solar zenith angle ($z$) as:

\begin{equation}
v(\beta,r_h,z)=v_0 \beta^\gamma r_h^\Gamma (\cos{z})^\varepsilon
\label{eq:velocity}
,\end{equation}

where $\gamma$ takes usually values of $\gamma\le$0.5 
\citep{1987A&A...171..327F}, $\Gamma$ is normally set to $\Gamma$=--0.5 \citep[e.g.][]{1951ApJ...113..464W,2000Icar..148...80R}, and  $v_0$ is a constant. The dependence of speeds on $z$ constitutes a new feature of the code that breaks the usual univocal correspondence between particle size and terminal velocity in previous dust tail codes \citep[see][for a discussion on the subject]{2011ApJ...732..104T}.  This dependence has been derived through three-dimensional dusty dynamical modeling by \cite{1997Icar..127..319C} and by \cite{2011ApJ...732..104T}, among others. 
Therefore, for the modeling of the dust shells and DLS features described in this paper (which, in all cases, only appear at perihelion or post-perihelion), we have adopted $\varepsilon$=0.5. However, we also note that \cite{2008Icar..193..572K} and \cite{2009AJ....137.4633K} adopted values of 
$\varepsilon$ of 1 or 0.25, depending on whether the velocity and dust production rate are proportional to insolation or to the surface temperature of a slowly rotating nucleus, respectively. Thus, there is actually a range of possible values of 
$\varepsilon$ that may be considered. Given this, we found it appropriate to model the pre-perihelion observation of comet C/1874 H1 (Coggia) by adopting 
$\varepsilon$=1, as we explain below (see Sect. \ref{sec:results}).

The principle of the Monte Carlo procedure is to compute the trajectories of a large number of particles of various sizes within the assumed size distribution, subject to the initial conditions described above, and to calculate their contribution to the brightness in each pixel of the image. The complete procedure is described in \cite{Moreno25a}.

To model the observed structures, we aimed to use a common pattern for all these comets as much as possible. It soon became apparent that these structures emerge only when very small dust particles are ejected, primarily within a few days before the observations. In all cases, we have assumed a size distribution limited by $r_{min}$=10$^{-8}$ m and $r_{max}$=10$^{-5}$ m, in other words between 0.01 and 10 micrometers in radius, with a power-law index of $\kappa$=--4, meaning that the distribution is entirely dominated in mass and brightness by very small particles. The presence of small particles or small aggregates in long-period comets is widely accepted, as they may, for example, explain the high levels of linear polarization observed in some cases, such as comets C/1995 O1 (Hale-Bopp),  C/1996 B2 (Hyakutake), and C/2001 Q4 \citep[NEAT;][]{2000Icar..145..203M,2009Icar..199..129L,
2009Icar..201..666G}, as well as in extrasolar comets \citep{2021NatCo..12.1797B}. Among the comets described in this paper, C/2020 F3 also exhibits a high degree of polarization \citep[about 25\% at a phase angle of 80$^\circ$, see][]{2021ATsir1648....1M}, while C/2023 A3 has shown a degree of polarization significantly higher than that of most short-period comets (Gray et al., in prep., 2025).

Regarding ejection velocities, we find that the above function (Eq. \ref{eq:velocity}) with $\gamma$=0.25 and $\varepsilon$=0.5, in other words

\begin{equation}
v(\beta,z)=v_0 \beta^{0.25} (\cos{z})^{0.5}
\label{eq:velocity2}
,\end{equation}

provides an adequate parametrization for modeling the observed tail features in all cases. The dependence on heliocentric distance has not been included due to the short time span required to model the structures. 

Since image scales are unknown for all the historic comets analyzed, we have adopted a reference value of 
$v_0$= 1 km s$^{-1}$ for all of them, which is, furthermore, the one found for comet C/2023 A3, as is shown in Sect. \ref{sec:results}. 


We emphasize that the purpose of the modeling is to identify the parameters that produce tail morphologies (dust shells and DLS) qualitatively resembling the observations. This approach is necessary because, in all cases, absolute brightness information is unavailable, and for only the two most recent objects, C/2020 F3 and C/2023 A3, the image scale is known. Regarding nucleus rotation parameters, there are estimates of the rotation period only for comets C/1858 L1  \citep[$P$=4.6 h,][]{1978Natur.273..134W}, C/1874 H1 \citep[$P$=9.2 h,][]{1981motc.conf..191W}, and C/2020 F3 \citep[$P\sim$7.5-7.8 h,][]  
{2020ATel13945....1D,2021MNRAS.506.6195M,2025Aravind}. However, since image scale information is unavailable for most objects, we cannot use the rotation period as a constraint in our models, except for comet C/2020 F3. For this comet, additional estimates of the nucleus spin axis orientation are available: \cite{2021MNRAS.506.6195M} report values of RA = 210$^\circ\pm$10$^\circ$ and DEC = +35$^\circ\pm$10$^\circ$, which correspond to an inclination of the spin axis to the orbital plane ($I$) and the argument of the subsolar meridian at perihelion ($\Phi$) of approximately $I\sim\Phi\sim149^\circ$. For the objects with unknown rotation period, we assumed, arbitrarily, $P$=12 h.

\begin{table*}
  \centering
  \caption{Model fitting parameters to retrieve the observed structures in each object.$^\star$
  }
    \label{tab:modelpar}
  \begin{tabular}{|l|c|c|r|r|r|r|r|}
    \hline
\multicolumn{1}{|c|}{Comet} & \multicolumn{1}{|c|}{Obs. date (UT)} & $r_h$ (au)& $\Delta$T(d)& 
 \multicolumn{1}{|c|}{$I^\circ$} & \multicolumn{1}{|c|}{$\Phi^\circ$}    & \multicolumn{1}{|c|}{$\lambda_{min}^\circ$} &\multicolumn{1}{|c|}{$\lambda_{max}^\circ$} 
\\
    \hline
C/1858 L1 (Donati) & 15-Oct-1858 & 0.663 & +15 & 165 & 190 & --60 & +20 \\
C/1874 H1 (Coggia) & 10-Jun-1874 & 0.902 &--29 &  \multicolumn{4}{|c|}{-- Sunward emission -- } \\
C/1874 H1 (Coggia) & 09-Jul-1874 & 0.676 & 0 &  90 & 180 & --55 & +55  \\
C/1881 K1 (Tebbutt) & 26-Jun-1881 & 0.757 & +9 &  90 & 160 & --10 & +85  \\
C/1901 G1 (Viscara) & 04-May-1901 & 0.391 & +11 &  90 & 80 & +10 & +40  \\
C/2020 F3 (NEOWISE) & 17-Jul-2020 & 0.495 & +14 &  160 & 170 & --50 & +50  \\
C/2023 A3 (Tsuchishan-ATLAS) & 13-Oct-2024 & 0.563 & +16 &  90 & 260 & --45 & 0 \\
  \hline

\multicolumn{8}{l}{$^\star$ $r_h$: heliocentric distance at observation; $\Delta$T: time to comet's perihelion; $I$: obliquity;}\\
\multicolumn{8}{l}{$\Phi$: argument of the subsolar meridian at perihelion; $\lambda_{min}$, $\lambda_{max}$: latitude limits of the active area.} \\

  \end{tabular}  
  \end{table*}

\section{Results} \label{sec:results}

The fitting procedure used to determine the model parameters is based on a trial-and-error approach. Figures 
\ref{fig:C1858L1} to \ref{fig:C2023A3} show the observations alongside the synthetic images for the parameters displayed in Table \ref{tab:modelpar}. In all cases, the synthetic images were oriented on the sky plane to approximately match those of the observed images, which do not necessarily follow the conventional “north up, east to the left” orientation. The position of the Sun is not shown in the plots, as it is always in the opposite direction to the dust shells. The most remarkable result regarding nucleus spin axes is that, to match the observed shell structures, the rotation axes of all nuclei should be located near the comet's orbital plane, at either $I\sim$90$^\circ$ or $I\sim$180 $^\circ$, while $\Phi$ varies from comet to comet. The tendency for nucleus spin axes to cluster near obliquities of either 90$^\circ$ or 180$^\circ$ was reported early on by \cite{1981AREPS...9..113S}, especially among short-period comets. It is primarily the result of dynamical evolution driven mainly by the effects of collisions in the early phases of the formation of the Solar System    \citep{2015A&A...582A..99I}, but also, to a lesser extent, long-term solar radiation influences such as the YORP thermal effect, which explain why the distribution of spin axes of small asteroids (smaller than $\sim$30 km) is clustered toward the ecliptic poles \citep{2011A&A...530A.134H}. These processes naturally lead to a statistical preference for higher obliquities over time. This high obliquity causes regions on the comet nucleus that were never illuminated before perihelion to become exposed to the Sun within a short time interval around perihelion. During this period, the activity strongly increases due to the small perihelion distance, which is smaller than 1 au for all the cases considered.

\begin{figure}
\centering
\centering\includegraphics[trim=6cm 0.5cm 7.5cm 2cm,clip,width=0.5\textwidth]{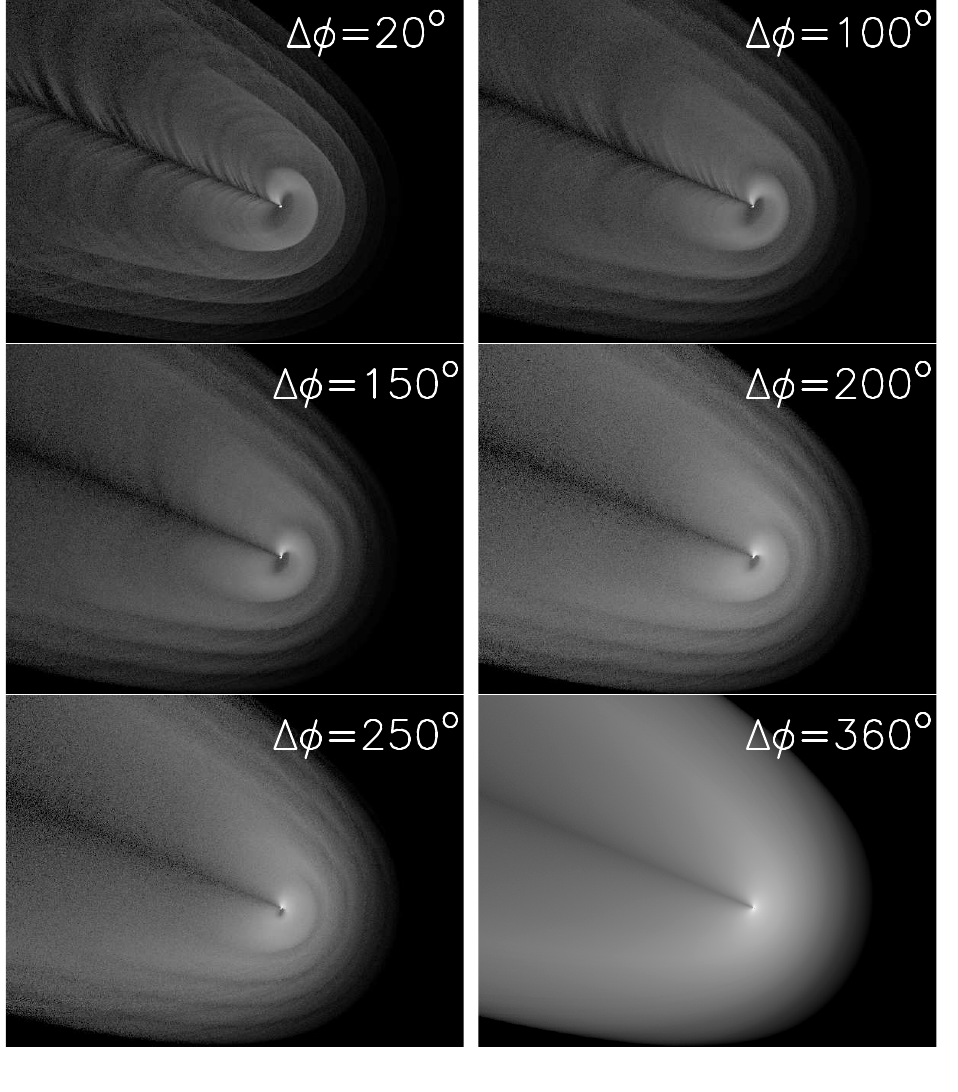}
\caption{Simulations of the image of C/2023 A3 on October 13, 2024, for several intervals of the active area longitude, as is indicated.}
\label{fig:lon-span}
\end{figure}

The longitude range of the active area ($\Delta\phi$) is the parameter that drives the spin-modulated activity, as the variation in the active surface area (which is proportional to the dust mass ejected) as a function of time will be periodic, leading to the formation of shell structures. We found that $\Delta\phi$ must generally be wide; otherwise, spurious structures in the tail displaying high-contrast brightness variations, which are not observed, would appear. However, it should not cover the full range from \(0^\circ\) to \(360^\circ\) either, because in that case, there would be no cyclic variation in the ejected dust mass, and the shells would not form. This is demonstrated in Fig. \ref{fig:lon-span}, where simulated images of the dust shells and DLS for comet C/2023 A3 are shown for different $\Delta\phi$ values. As is seen, the spurious structures vanish for $\Delta\phi \gtrsim 200^\circ$. We finally adopted a value of $\Delta\phi$=250$^\circ$, which we found to be adequate for all the cases analyzed. 

In Table \ref{tab:modelpar}, the selected range of latitudes set to generate  the synthetic images is indicated. Since there are no constraints on the absolute brightness for any of the images, we cannot determine the latitude range (which will be proportional to the ejected mass) more precisely, so the values of $\lambda_{min}$ and $\lambda_{max}$ are only loosely constrained, the real latitude extent might be somewhat different. 

\begin{figure}
\centering
\centering\includegraphics[trim=1cm 0.5cm 4cm 3cm,clip,width=0.5\textwidth]{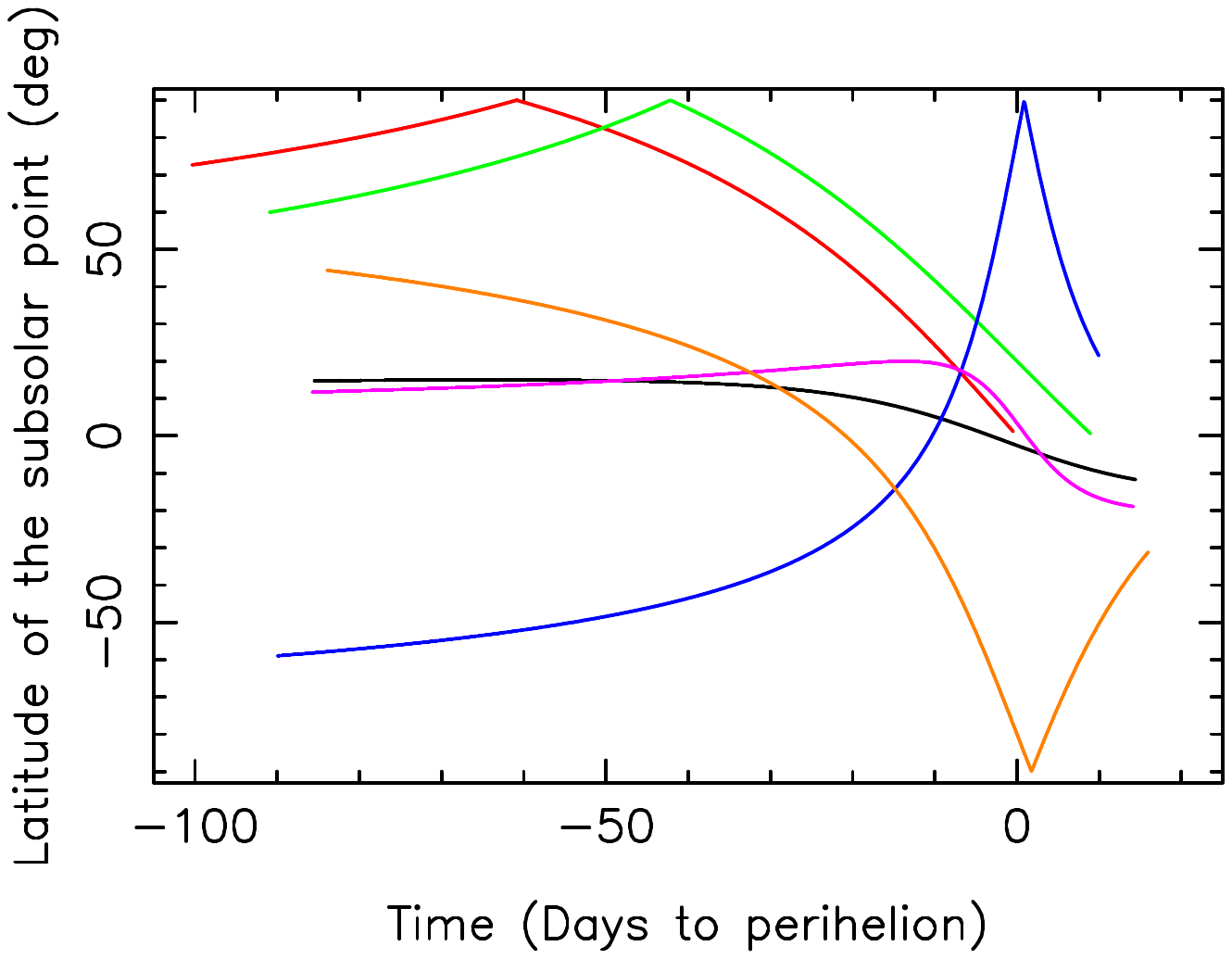}
\caption{Latitude of the subsolar point as a function of time for the objects listed in Tables \ref{tab:logobs} with the rotational parameters displayed in Table \ref{tab:modelpar}. The different colors correspond to the distinct objects as follows: black, C/1858 L1; red, C/1874 H1; green, C/1881 K1; blue, C/1901 G1; purple, C/2020 F3; and brown, C/2023 A3. }
\label{fig:subsolar}
\end{figure}

\begin{figure}
\begin{tabular}{c}
\includegraphics[trim=1cm 1.5cm 3cm 3cm,clip,width=0.5\textwidth]{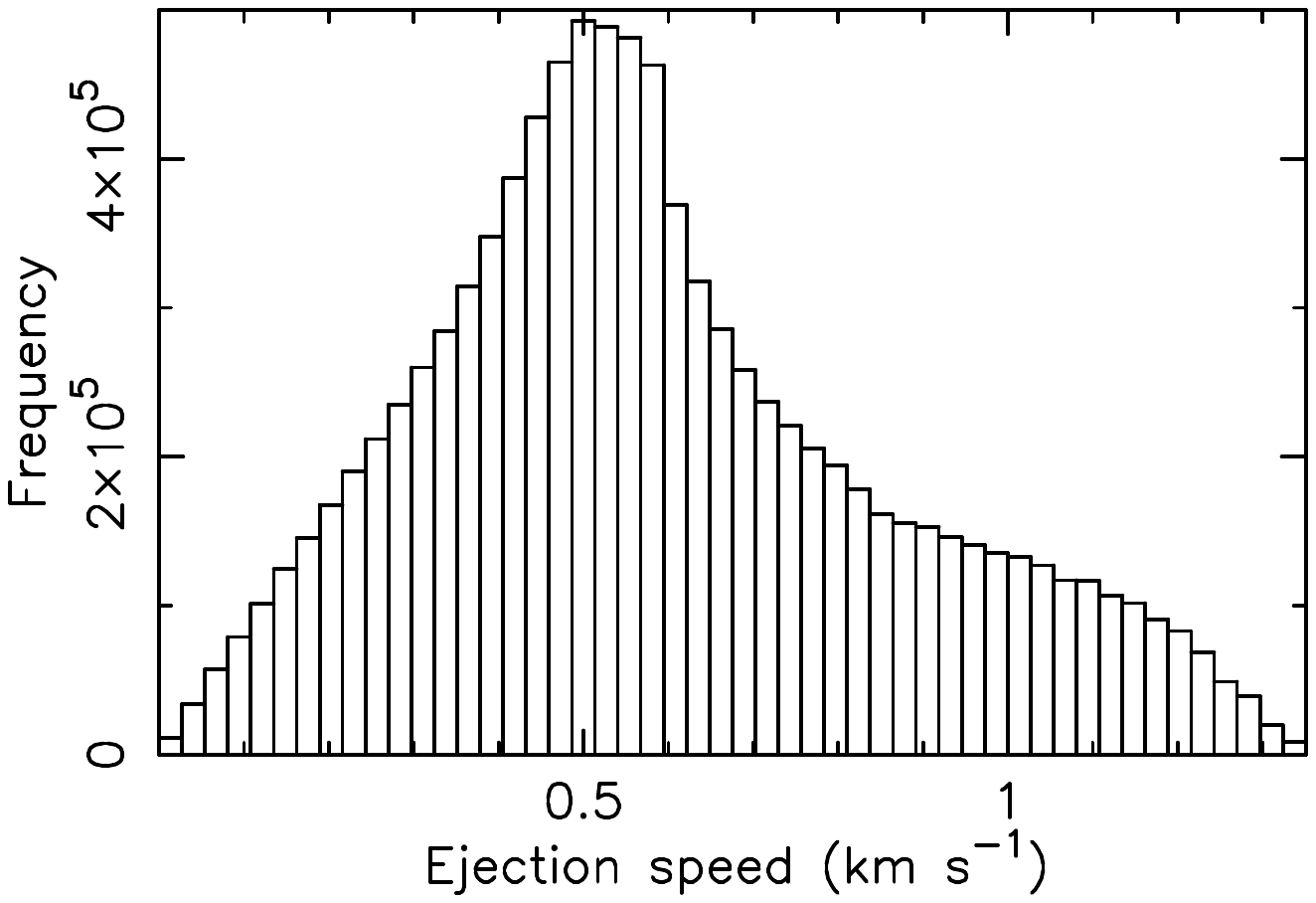} \\  
\includegraphics[trim=1cm 1.5cm 3cm 3cm,clip,width=0.5\textwidth]{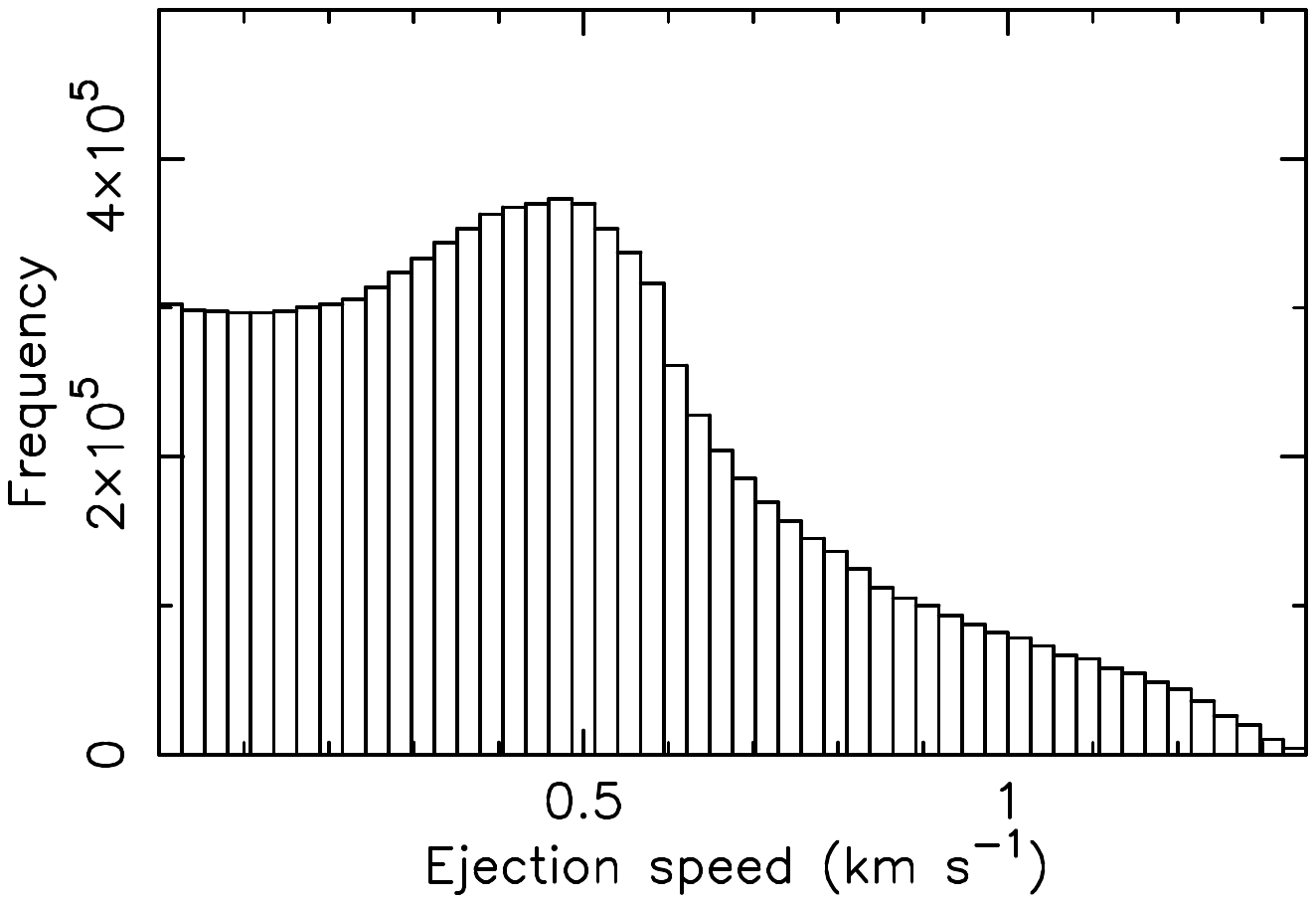}
\\
\end{tabular}
\caption{Histograms of the speed distribution of the particles ejected from the nucleus of comet C/1874 H1 (Coggia). Upper panel: Assuming $v\propto  \sqrt{\cos {z}}$. Lower panel: Assuming $v\propto \cos{z}$.}
\label{fig:histogram_vel}
\end{figure}


\begin{figure}
\centering
\centering\includegraphics[angle=-90,trim=2cm 2cm 5cm 2cm,clip,width=\columnwidth]{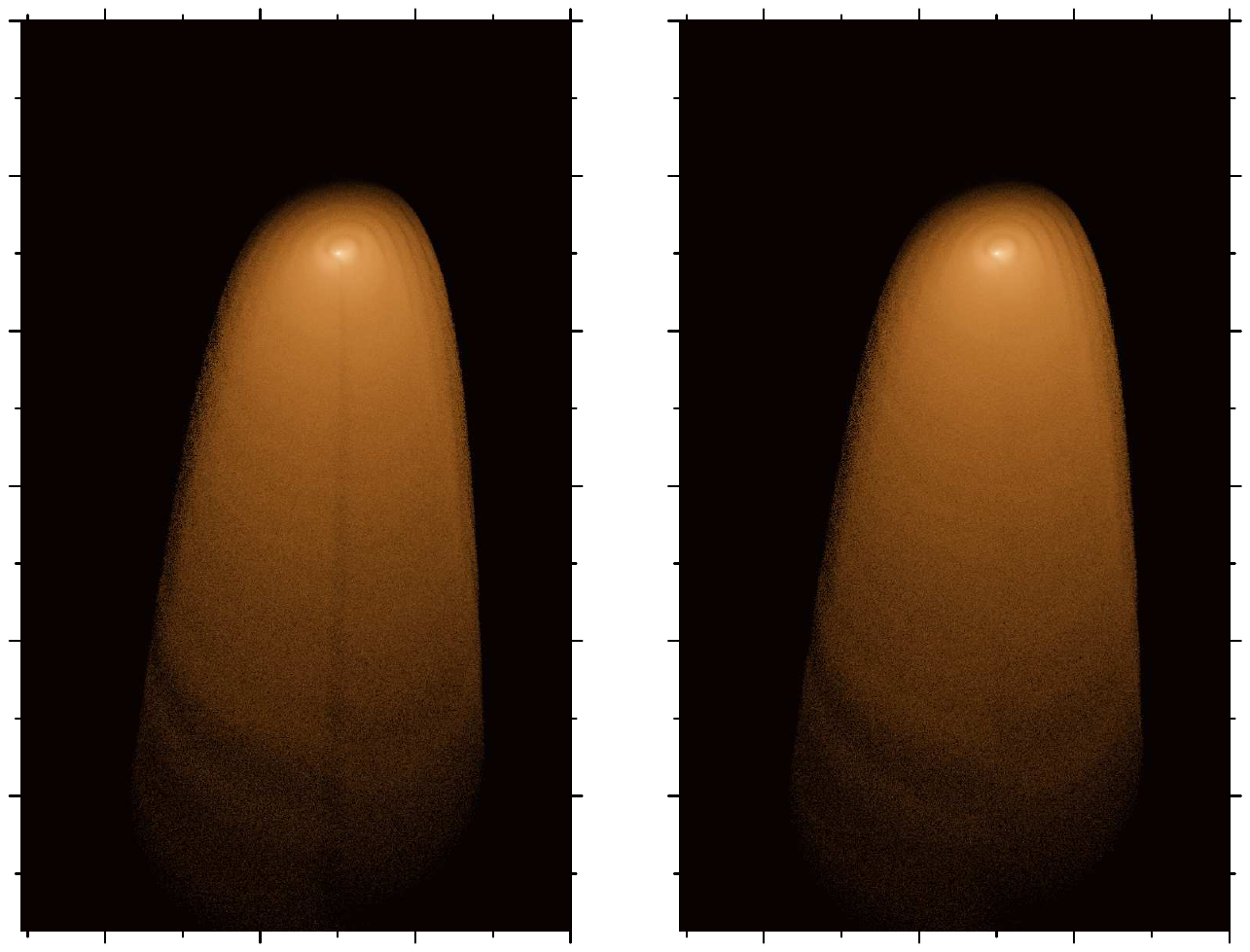}
\caption{Top panel: Synthetic image to model the dust tail of comet C/1881 K1 (Tebbutt), with $[\lambda_{min}, \lambda_{max}]$=$[-10^\circ,+85^\circ]$. Bottom panel: Same model as in the top panel but for $[\lambda_{min}, \lambda_{max}]$=$[-10^\circ,+90^\circ]$, where the DLS vanish.}
\label{fig:DLS_noDLS}
\end{figure}

While the longitude extent is mostly responsible for the shell generation, the DLS is a bit more complex to explain. We observed that in all cases the dependence of the ejection speeds on some function of $\cos z$ is playing a significant role. To illustrate this, we take the example of comet C/1874 H1 (Coggia). For this object, we modeled the perihelion image in the same way as all the other objects; that is, with an emission speed proportional to $(\cos z)^{0.5}$. However, the pre-perihelion image reveals a structure that not only lacks the shells but, more importantly, exhibits a bright linear feature near the tail axis instead of the DLS observed in the perihelion image. We emphasize that we interpret this bright stripe as a dust feature. However, we note that it could also be due to an ion stream, but owing to the limited spatial resolution of the drawing, we cannot clearly establish its origin.

This bright feature might be explained by an excess of low-speed particles accumulating near the tail axis. One way to model both this excess emission and the absence of shells could be to adopt a sunward ejection model instead of the single active area model at perihelion, and to modify the dependence of emission speeds from $(\cos z)^{0.5}$ to a simple proportionality with $\cos z$. This is qualitatively in agreement with the findings of \cite{2011ApJ...732..104T}, who obtained, for $r$=1 $\mu$m particles, an increase in the ratio of speeds at 0$^\circ$ solar zenith angle to speeds at 90$^\circ$ solar zenith angle with increasing heliocentric distance (the dependence of speeds with $z$ using $v \propto \cos z$ is steeper than with $v\propto (\cos z)^{0.5}$).

These two different speeds distributions are depicted in Fig. \ref{fig:histogram_vel}. With these distinct distributions, we obtained the remarkably different synthetic images of comet C/1874 H1 (Coggia) shown in Fig. \ref{fig:C1874H1}, which qualitatively compare very well with the observed ones. However, it is not only the speeds (magnitudes) of the velocity vectors that play a role but also their directions. To show this, we refer to the simulations of the dust tail of comet C/1881 K1 (Fig. \ref{fig:C1881K1}). This figure shows the qualitative agreement with the observed image of this comet, reproducing the shells and DLS, including the slight asymmetry observed toward the upper right corner of the plots. For this object, we set $[\lambda_{min}, \lambda_{max}]$=$[-10^\circ,+85^\circ]$ (Table \ref{tab:modelpar}). However, by setting the upper limit to $\lambda_{max}=+90^\circ$, we see that the shells are still there, but the DLS vanishes (see Fig. \ref{fig:DLS_noDLS}), demonstrating the fact that the velocity field vector is also playing a fundamental role in the DLS formation.

A similar modeling procedure as the one just described for the above objects was applied to comets C/2020 F3 and C/2023 A3. For these two objects,  we have information on the image scale, and for the former, also on the rotational period. This allowed us to fit the model parameters more accurately. In the case of C/2020 F3, the image scale is 0.12 arcsec px$^{-1}$,  and the best-fitting velocity parameter was found at $v_0$=0.5 km s$^{-1}$, and the rotational parameters $I$ and $\Phi$ at 160$^\circ$ and 170$^\circ$, respectively, not far from the estimates by  \cite{2021MNRAS.506.6195M} of  $I\sim\Phi\sim149^\circ$. In Fig. \ref{fig:C2020F3}, we compare the observation with the simulated image, in which 
the red dots indicate the position of the first two dark spaces between dust shells along the tail axis in both images for comparison, showing a good agreement, as well as in the DLS appearance.

The case of C/2023 A3 is displayed in Fig. \ref{fig:C2023A3}. In this case, the image scale is known (1.5 arcsec px$^{-1}$), and the simulation was performed assuming arbitrarily a rotation period of 12 h, as is stated in Sect. \ref{sec:model}.  The best fit velocity  parameter was found at $v_0$=1.0 km s$^{-1}$. As Fig. \ref{fig:C2023A3} shows, the shell separation is well matched, as is revealed by the position of the red dots in both images, which show, as before, the first two dark spaces between shells. The DLS is also well reproduced.  

As is indicated in Sect. \ref{sec:model}, in our model the shells and DLSs are driven by a transient activity superimposed on the background cometary activity, triggered by the exposure of fresh nucleus areas near the comet's perihelion. To illustrate this, we refer to available observations of C/2023 A3 on the internet; for example, on AstroBin.\footnote{\url{https://es.welcome.astrobin.com/}} These images show that the DLS (and possibly the shells, although this cannot be confirmed) were visible from October 13 to October 21, 2024, but were no longer seen on November 7. Assuming transient activity lasting from October 10 to October 21, we simulated the November 7 image, in which neither the shells nor the DLS are visible -- only a diffuse tail with brightness levels at least four magnitudes lower than in the October 13 image, as most particles have left the field of view due to strong radiation pressure on the smaller particles in the distribution.

\section{Conclusions} \label{conclusions}

We have shown that the shells and the DLSs that appear along the tail axis of several historical and recent long-period comets can be modeled by transient activity lasting a few days close to the perihelion time, superimposed on the comet background emission. The modeled dust particles share common dust physical parameters, namely the same refractive indices and size distribution, characterized by very small particles. On the other hand, all of those objects exhibit high inclinations of their spin axes, favoring a rapid increase in the solar flux over freshly exposed areas on the nucleus surface near perihelion. The dust shells appear because of the spin modulated activity as a consequence of dust 
ejected from a single active area exposed periodically to solar radiation as the nucleus rotates, a mechanism that was established many years ago by F. Whipple \citep[e.g.][]{1978Natur.273..134W}. On the other hand,  the DLSs along the tail axes, which appear to be associated with those halo dust patterns, originate from the specific distribution of ejection velocities. This distribution might be modeled as the velocity dependence on the square root of the cosine of the solar zenith angle at the particle emission point, supporting the findings from hydrodynamical modeling. In addition, 
the presence or absence of the DLS in the images depends not only on the speed magnitude but also on the specific latitude extent, which defines the velocity vector field. Finally, although the analysis has primarily focused on long-period comets, we would like to mention that structures similar to the DLSs are also sometimes observed in other comets, such as the “dark lane” reported in the Halley-type comet 12P/Pons-Brooks after an outburst \citep{2025MNRAS.tmp..295G}, which might have formed in a similar manner.

\begin{acknowledgements}
    
We are indebted to an anonymous referee for their very useful comments and suggestions.

FM acknowledges financial supports from grants PID2021-123370OB-I00, and from the Severo Ochoa grant CEX2021-001131-S funded by MCIU/AEI/ 10.13039/501100011033. 

EJ is Director of Research at the Belgian FNRS.

Jos\'e Carrillo of amateur astronomical association \texttt{Cometas\_Obs}, and Damian Peach are gratefully acknowledged for sharing the C/2023 A3 and C/2020 F3 images displayed in Figures \ref{fig:C2023A3-1}, \ref{fig:C2020F3}, and \ref{fig:C2023A3}. 

Colin Snodgrass is gratefully acknowledged for providing useful comments on the manuscript. 

The \texttt{FORTRAN} code \texttt{COMTAILS} makes use of the JPL Horizons on-line ephemeris system.

\end{acknowledgements}

\bibliographystyle{aa} 
\bibliography{aa53986-25corr} 

\end{document}